\documentclass{journaleducation}
\begin{document}
\title{Democratising Optical Orbital Angular Momentum: a Set of Cost-Effective Tools 
}
\vspace{5mm} 
{\fontsize{10}{12} 
Natasha Bierrum$^1$*, Lyuxuan Chen$^1$, Ananya Kudaloor$^2$, Lok Kan Wan$^2$, Shupeng Yang$^2$, Yancen Hou$^2$, Xiwen Dong$^2$, Muskan Tuli$^2$, Richard Taylor$^{1,3}$, Petros Androvitsaneas$^1$, Carrie Weidner$^1$, Edmund Harbord$^1$ }\\

\onehalfspacing
{\fontsize{8}{10}\selectfont \raggedright $^1$Quantum Engineering Technology Labs, H. H. Wills Physics Laboratory and School of Electrical, Electronic and Mechanical Engineering, University of Bristol, BS8 1FD, United Kingdom\\
$^2$School of Electrical, Electronic and Mechanical Engineering, University of Bristol, BS8 1TR, United Kingdom\\
$^3$Vector Photonics Limited, Building 4.05, West of Scotland Science Park, Kelvin Campus, 2317 Maryhill Rd, Glasgow, G20 0SP, United Kingdom\\
$^*$Author to whom any correspondence should be addressed.}
\singlespacing
{\fontsize{8}{10}\selectfont \raggedright {\bfseries E-mail:} n.bierrum@bristol.ac.uk}

% \keywords{Optical orbital angular momentum, quantum technology, pedagogy, demonstration}

\section*{Abstract}
\normalsize Classical and quantum optical communication has gained popularity and momentum in recent years, with growing investment and innovation in quantum technologies. However, the main teaching method in the education of quantum mechanics include mathematically-intensive derivations or abstract analogies for the complex systems. We propose a "poor man's" spatial light modulator experiment that is an engaging and interactive learning aid for teaching quantum mechanics and optical orbital angular momentum. Fork diffraction gratings were created on photographic slide film by outsourcing to an external company, and so the gratings were easy and cheap to produce.
A simple setup with a fork diffraction grating and a laser pointer successfully produces vortex beams that possess orbital angular momentum, allowing for orbital angular momentum to be easily observed and investigated in a teaching environment. How the tools can be used effectively to enhance learning is discussed, either as a demonstration or as an investigative scientific learning environment activity. 

\section{Introduction}
Quantum technologies are changing the world, and it is vital to teach quantum physics concepts to school children and undergraduates \cite{seifollahi_preparing_2025,fox_preparing_2020} to facilitate a quantum ready workforce \cite{greinert_future_2023,aiello_achieving_2021}.
While there are a growing number of virtual labs, visualisers, and "quantum games" that have entered the pedagogic arena, there is a dearth of simple, "hands-on", and economically viable demonstrations for these quantum apprentices. This paper seeks to report the basis of such an activity.
One of the most conceptually challenging properties in quantum photonics is orbital angular momentum (OAM) - since its inception and experimental demonstration in 1992, it has been of great interest to quantum technologists \cite{allen_orbital_1992,chen_orbital_2020}. 
Suitably prepared light can carry this OAM, giving rise to a theoretically unbounded number of orthogonal modes, OAM allows for efficient transfer of information making it attractive for optical communication \cite{willner_optical_2015, gibson_free-space_2004}, quantum cryptography \cite{mirhosseini_high-dimensional_2015,larocque_generalized_2017}, and quantum key distribution \cite{malik_influence_2012}. \\
Light carrying OAM can be readily produced with a laser and diffractive optics, and so has the potential to be utilised as an active learning demonstration in a lecture or a laboratory activity to support students' learning and engagement \cite{munoz-losa_impact_2025}. 
Diffraction is a key concept taught at school level and the production of OAM with diffractive optics easily lends itself to an activity for introducing quantum mechanics to school students. 
The equipment and tools for the experiment therefore need to be readily available, low-cost, and suitable for use in teaching environments.
% as a demonstration in a lecture or laboratory experiment. 
Diffractive optics based on amplitude diffraction gratings have been printed on transparent sheets for demonstrating diffraction for teaching \cite{van_hook_inquiry_2007}, and for producing and identifying OAM beams in research laboratories \cite{vaziri_simple_2024,dai_measuring_2015}. 
However, pixelation of the grating and an upper limit on the lines per mm that can be created with an inkjet printer restrict the designs achievable. An alternative approach is successfully demonstrated in this paper, where diffraction gratings are created on $35 \ $mm photographic film.
The process of image setting on photographic film has a resolution of up to 4800 dpi, compared to 1200 dpi for a typical inkjet printer \cite{lee_fabrication_2010}. The production of the diffraction gratings on photographic film was outsourced to an external company and was still found be to be a cost-effective method.
Diffraction gratings produced on photographic film have not been previously shown or tested to produce OAM beams, but they have been shown to create Young's double slits \cite{velentzas_teaching_2014,lee_fabrication_2010}. This work shows that diffraction gratings on photographic film can be used to produce beams carrying OAM and that it is advantageous for researchers as well as educators.
In the following sections, the process for creating the diffraction gratings on photographic negative film and positive film slides is outlined and analysed. Finally, we discuss how the diffraction gratings on film can be used in undergraduate and school teaching in laboratory activities involving students and as a demonstration in a lecture or classroom.

% \section{Theoretical background}
\subsection{Optical orbital angular momentum theory}
\begin{figure}
    \centering
    \includegraphics[width=0.72\linewidth]{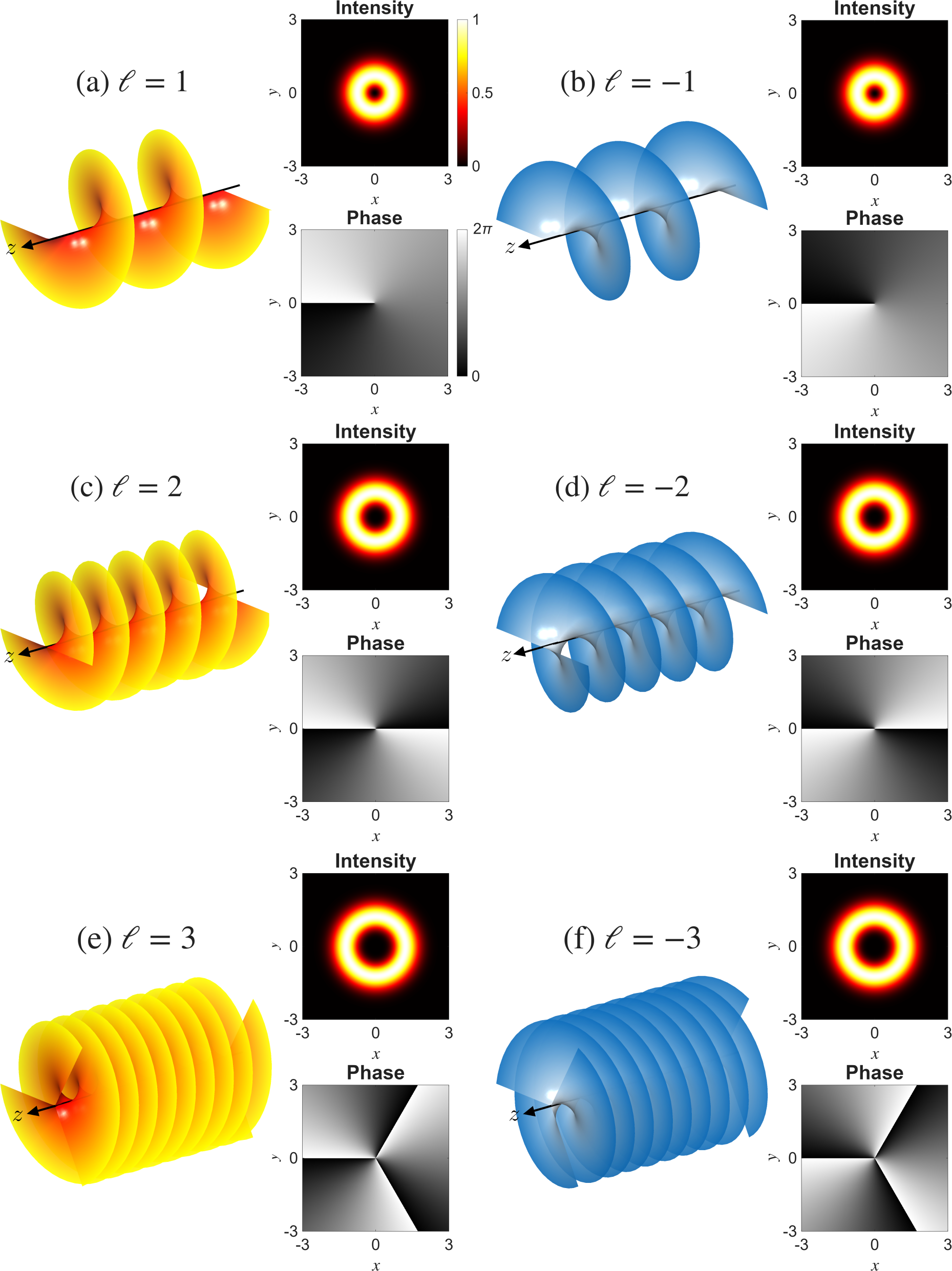}
    \caption{Plots of the helical wavefront, normalised intensity distribution in arbitrary units, and phase profile in radians for Laguerre-Gaussian beams of selected topological charge values $\ell$, and $p=0$. (a) and (b) show the same magnitude of $|\ell|=1$, hence the intensity plots appear identical, but the phase profile and  handedness of the wavefront are reversed. The same is true for (c) and (d) with $|\ell|=2$, and (e) and (f) with $|\ell|=3$.
    The colour bars in (a) apply to all subplots.}
    \label{fig:oamfig}
\end{figure}

Light carrying OAM has a distinct helical wavefront and a spiral phase variation in the transverse plane of the beam.
Each photon carries $\hbar \ell$ of OAM, where $\ell$ is the number of intertwining helical wavefronts and $\hbar$ is the reduced Planck's constant, showing that OAM is a distinctly different mechanism 
to the spin angular momentum carried by circularly polarised light \cite{allen_orbital_1992}. 
The Laguerre-Gaussian ($LG$) modes represent the profile of OAM beams, and they are described as
\begin{equation}
    LG_{p}^\ell (r, \phi, z) = A_{p}^\ell (r,z) \exp \left( i \ell \phi \right),
    \label{eq:LG}
\end{equation}
expressed here in the cylindrical coordinate system with radial $r=\sqrt{x^2+y^2}$, azimuthal $\phi$, and vertical $z$ coordinates \cite{yao_orbital_2011}. The amplitude distribution, $A_p^l(r,z)$, depends on the azimuth wavefront number $\ell$ and radial wavefront number $p$ and is explicitly expressed in the cylindrical coordinate system in Appendix \ref{s:LGmode}. %something about the exp term
For the OAM beams studied in this paper and generally used in optical communication applications \cite{gibson_free-space_2004}, $p=0$ and $\ell$ is named the topological charge of the OAM beam \cite{dai_measuring_2015}. The topological charge is an integer value that can be positive or negative, and the wavefront, intensity and phase for OAM beams of charge $\ell = \pm 1, \pm 2, \pm 3$ are shown in Figure \ref{fig:oamfig}.\\
Due to the phase singularity at the origin, OAM beams have a vortex in the centre of the beam where the intensity is zero. The diameter of the ring scales as $D \sqrt{\ell + 1}$ where $D$ is the Gaussian beam waist at position $z$ when $\ell=0$ \cite{dai_measuring_2015}, and is visible in the simulated intensity plots in Figure \ref{fig:oamfig}. \\
\newpage
\subsection{The production of orbital angular momentum beams}
\begin{figure}[h!]
    \centering
    \includegraphics[width=0.47\linewidth]{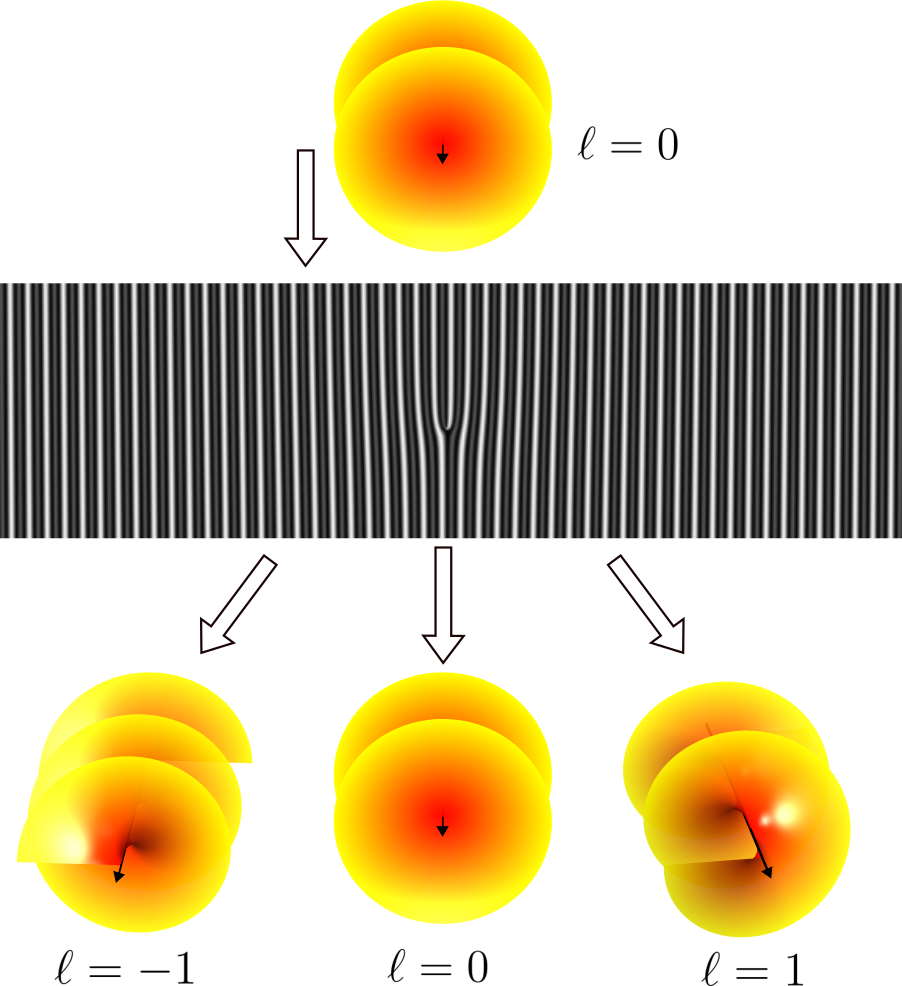}
    \caption{Illustration demonstrating the production of OAM beams with a hologram in the first and zeroth diffraction orders. A fundamental Gaussian mode, equivalent to $\ell=0$, is incident on a fork diffraction grating of charge $\ell=1$, creating OAM beams of topological charge $\ell = 1, -1$ in the first diffraction orders (positive and negative respectively), and the original Gaussian beam is observed in the zeroth diffraction order.
   }
    \label{fig:oamfromfork}
\end{figure}

There are multiple different methods of creating OAM beams, with advantages and disadvantages for each application, including cylindrical lenses \cite{allen_orbital_1992}, spiral phase plates \cite{beijersbergen_helical-wavefront_1994,oemrawsingh_production_2004}, and computer generated holograms \cite{heckenberg_generation_1992,stoyanov_far_2015}.
The latter of these methods is particularly well-suited for educational applications, since computer generated holograms create the desired phase profiles without the use of complex refractive optics.
A type of hologram that creates OAM beams in the far field is a diffraction grating with a defect in the centre, where a fringe splits in two or more, making a fork shape \cite{yao_orbital_2011}. 
The topological charge of the output beams is determined by the number of fringes that the central fringe splits into, hence Figure \ref{fig:oamfromfork} demonstrates a fork diffraction grating of charge $\ell = 1$. The result of an incoming optical beam incident is also depicted in the figure. \\
 \\
When a fundamental Gaussian mode TEM$_{00}$, equivalent to $\ell=0$, propagates through the region of the fork, OAM beams with topological charges $\ell$ and $-\ell$ are produced in the first diffraction orders $m=1$ and $m=-1$ respectively \cite{heckenberg_generation_1992,yao_orbital_2011}. 
Additionally, the input fundamental Gaussian mode is still visible in the zeroth diffraction order, showing that fork diffraction gratings are only partially efficient at creating desired OAM beams.
In fact, the diffraction grating can produce OAM beams in any diffraction order, following the relation $\ell_{m} = \ell_i + mq$, where $q$ is the charge of the hologram, $\ell_m$ is the topological charge of the OAM beam in the corresponding order of diffraction, $\ell_i$ is the topological charge of the input beam, and $m$ is the diffraction order \cite{stoyanov_far_2015}.
The relative intensities in the diffraction orders will depend on the hologram properties and can be optimised for certain orders of diffraction \cite{gibson_free-space_2004}. 
For example, with a fork hologram of charge $q=2$ and an input beam of $\ell=-2$, the fundamental Gaussian beam TEM$_{00}$ or $\ell=0$ can be recovered in the first diffraction order, $m=1$. \\
In research applications, computer generated holograms are typically implemented with a spatial light modulator (SLM): a reconfigurable, pixelated liquid crystal device with variable phase control \cite{yang_review_2023}.
The holograms are directly uploaded from a computer onto the SLM and so can produce a range of diffraction patterns at a variety of different wavelengths \cite{gibson_free-space_2004,zheng_measuring_2017}. However, an SLM is a highly specialised optical component that is expensive and not suitable for use outside of an optics laboratory setting.
In contrast, the fork diffraction grating on photographic film is a fixed amplitude hologram that produces one diffraction pattern optimised at one specific wavelength, but it is more accessible and suitable for teaching. 

\section{Experiment details and results}
% In this section, the method for creating the initial diffraction gratings on photographic film is discussed in Section \ref{ss:cgh} and the gratings are evaluated in Section \ref{ss:analysis}.
% Their efficiency for producing OAM beams is tested experimentally in Section \ref{ss:experiment}.
\subsection{Computer generated holograms}
\label{ss:cgh}
\begin{figure}[h!]
    \centering
    \includegraphics[width=0.77\linewidth]{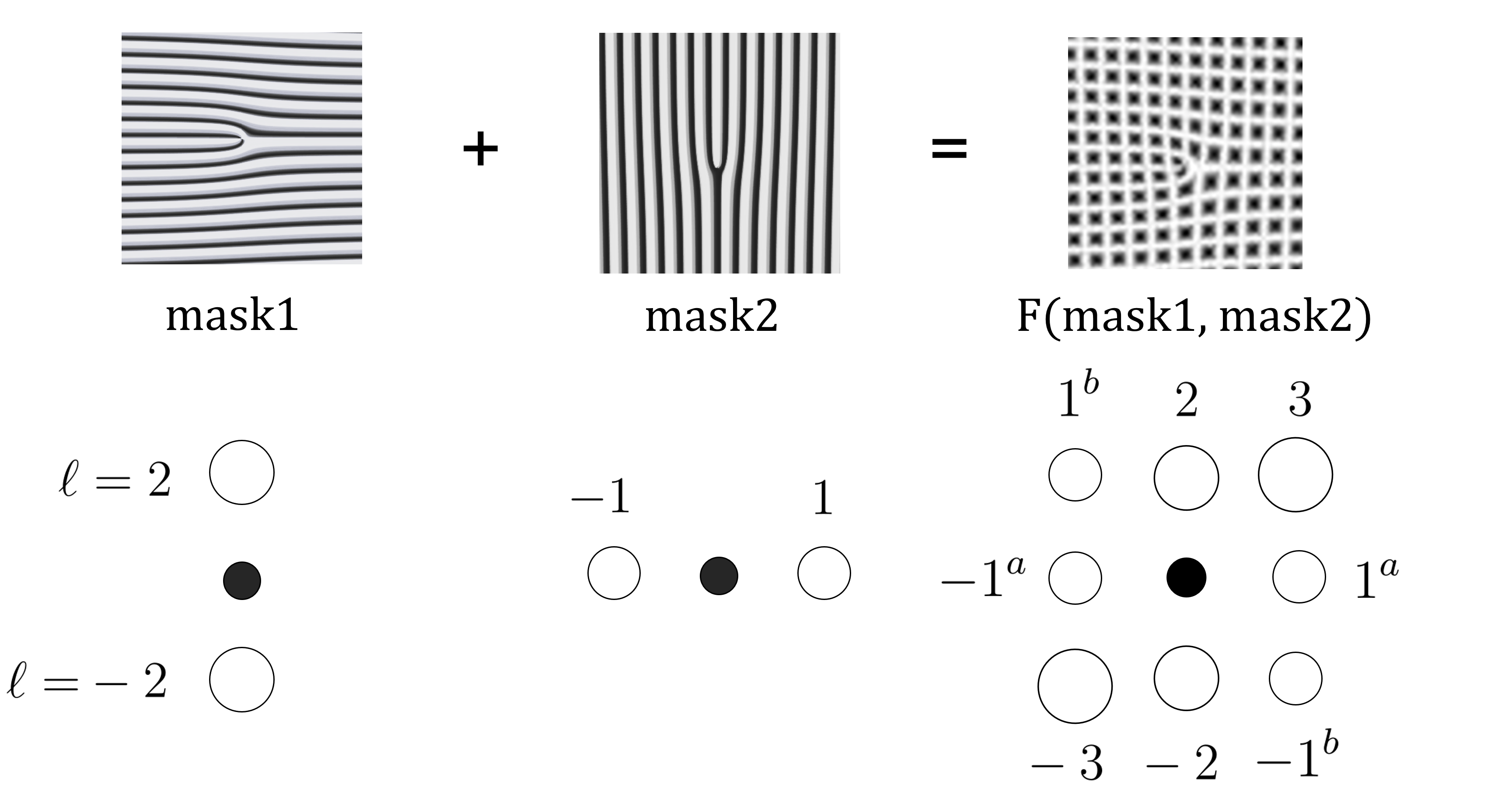} 
    \caption{Design and output of the combined fork diffraction grating of mask 1 and mask 2 for an incoming Gaussian beam. Mask 1 has a charge of $\ell=2$ and mask 2 has a charge of $\ell=1$, and the combined diffraction grating creating an array of $3 \times 3$ beams of different topological charges $\ell=-3 \ldots 3$. There are multiple beams of topological charge $\ell=1$ and $\ell=-1$, which have been labelled $a$ and $b$ for completeness, and the unlabelled black filled circles are the Gaussian modes $\ell=0$. Figure inspired by \cite{gibson_free-space_2004}.}
    \label{fig:combined_grating}
\end{figure}

\begin{figure}[h]
    \centering
    \includegraphics[width=0.6\linewidth]{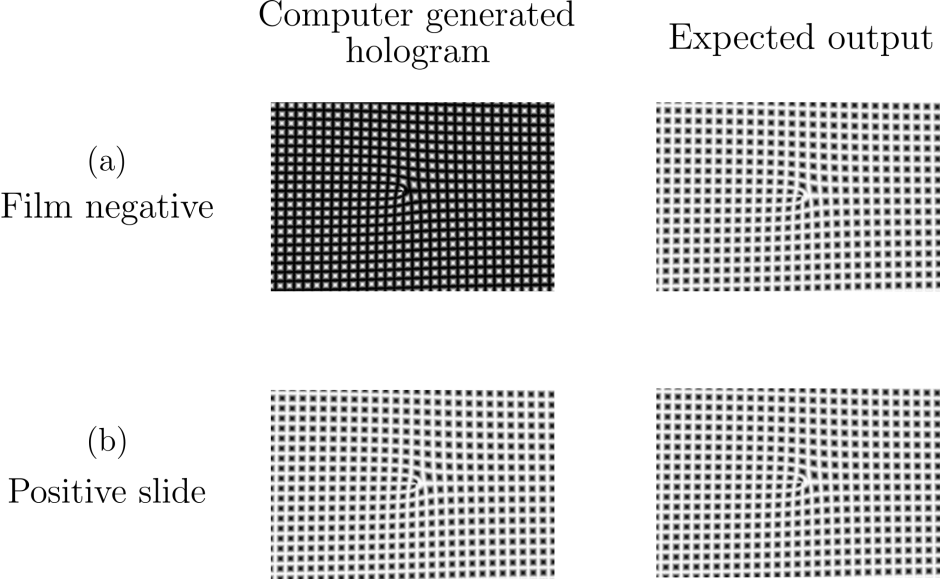}
    \caption{The computer generated hologram design and the expected output on photographic film for (a) negatives and (b) positive slides. The dark regions correspond to areas of high optical density that block the transmission, while the clear areas allow light to pass through.} 
    \label{fig:positivenegative}
\end{figure}

The computer generated holograms were created in MATLAB as previously reported \cite{vaziri_simple_2024}, and suitable diffraction grating periods were considered. 
% and the code including the equations for creating the fork diffraction gratings can be found in the repository.
A laser diode with a wavelength of 650 nm was used for testing, which is a typical red laser pointer wavelength. A range of periods from 50 \textmu m to 200 \textmu m and charges of $\ell=1$ to $\ell=6$ were designed specifically for this wavelength.
In addition, a more complicated design was created, the combination of two fork diffraction gratings placed at 90$^{\circ}$ to each other \cite{ii_increasing_2004}. This hologram produces a $3 \times 3$ array of beams with various topological charges, as shown in Figure \ref{fig:combined_grating}. 
\\
In order to observe the full array of topological charges, an optimisation function F(mask1, mask2) is applied to create the combined diffraction array. This function was found from simulations to produce equal intensity in each of the beams in the array \cite{ii_increasing_2004}. 
The type of photographic film and requirements for production also had to be considered for each of the holograms.
The company making the diffraction gratings on 35 mm film, \textit{Digital Slides} \cite{digital_2025}, requested that the amplitude of the grating was converted to an 8-bit JPEG image with resolution 4096 pixels $\times$ 2732 pixels. 
Depending on the type of film used to produce the diffraction gratings, either slides or negatives, the computer generated hologram was altered. 
Since the negative film process reverses the contrast of the image, the hologram is inverted beforehand to produce the same output as on the positive image slide film. If the hologram was not inverted, the grating would still produce OAM beams, but the film would mostly block the beam transmission. 
The difference in the process is summarised in Figure \ref{fig:positivenegative}.
\subsection{Analysis of fork diffraction gratings on film}
\label{ss:analysis}
\begin{figure}[h!]
    \centering
    \includegraphics[width=0.33\linewidth]{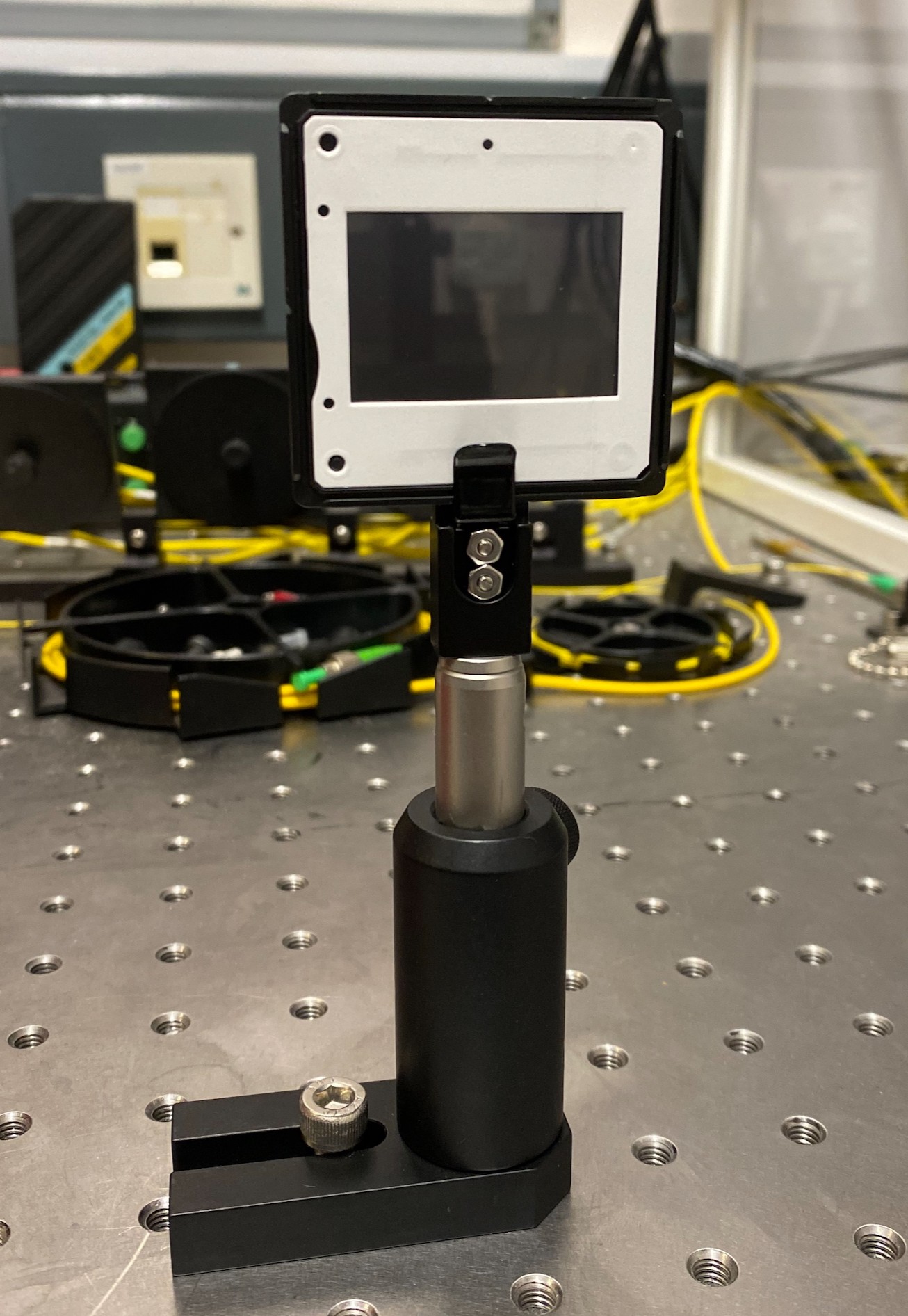}
    \caption{Diffraction grating on photographic film in plastic mounting holder. }
    \label{fig:slideholder}
\end{figure}

Both colour positive slides and black and white negatives were made with various fork diffraction grating designs.
The photographic films were ordered online from \textit{Digital Slides} \cite{digital_2025} and although the images were greyscale, the company requested RGB images for production of both black and white negatives and colour slides. 
The fork diffraction gratings cost £3.37 per colour slide in a plastic mounting holder and £2.50 per black and white negative. However, there is a larger minimum order of 12 units for black and white negatives, and the negatives are delivered as a strip, so there is extra cost for plastic holders.
The plastic mounting holders were used to hold the diffraction grating in a clamp (Thorlabs compact dual filter holder DH1), as shown in Figure \ref{fig:slideholder}.
\\

\begin{figure}[h!]
\centering
\begin{subfigure}{0.48\textwidth}
    \centering
    \includegraphics[width=\textwidth]{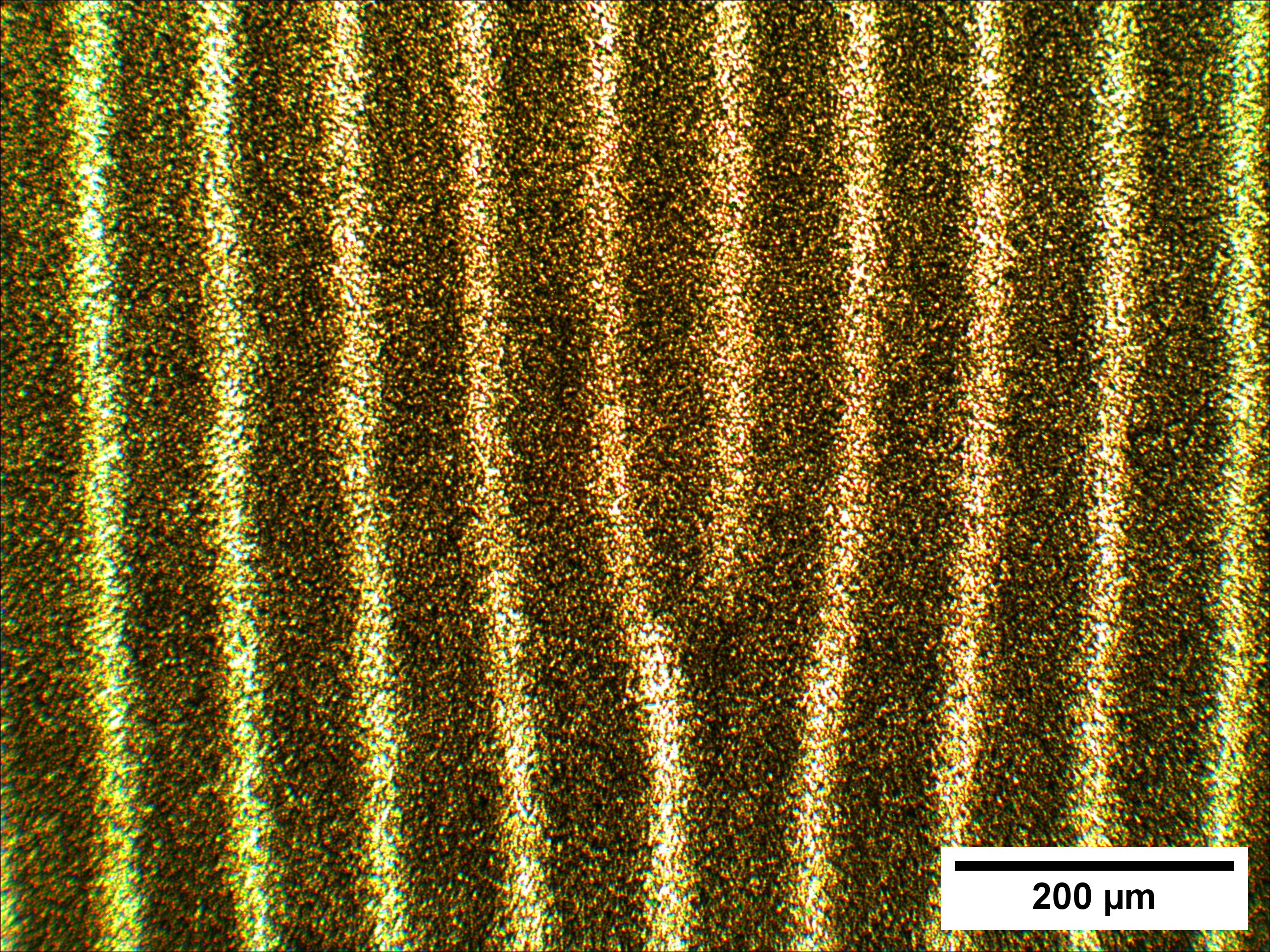}
    \caption{}
    \label{fig:first}
\end{subfigure}
\hfill
\begin{subfigure}{0.48\textwidth}
    \includegraphics[width=\textwidth]{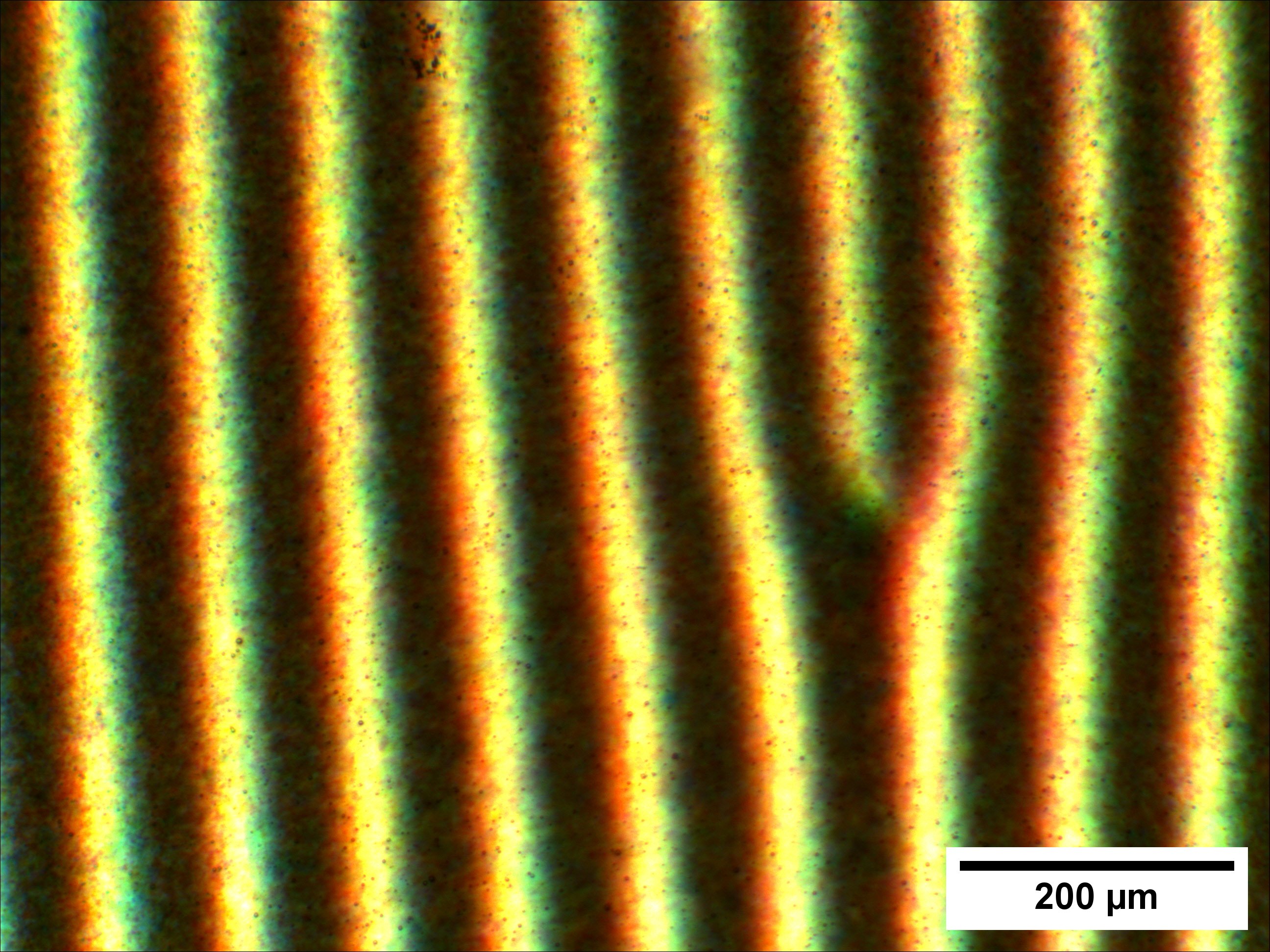}
    \caption{}
    \label{fig:second}
\end{subfigure}
\caption{Optical microscope images of the photographic film fork diffraction gratings on (a) black and white negative and (b) colour slide film.}
\label{fig:gratings}
\end{figure}

\noindent In this study, the slide film is Kodak Ektachrome E100 and the negative film is Ilford Pan F Plus ISO 50, which are both fine grain films. 
The process of creating the hologram is the same for both films, and the films are written with a cathode ray tube (CRT) film recorder, according to the company's procedure. The CRT film recorder creates a smoothed image on film without the pixelated structure of the original image.
%a Cathode Ray Tube (CRT) film recorder. The intensity of the beam was modulated according to the pixel intensity in each RGB channel and exposed with the corrsponding colour filter to produce the high quality colour. 
% The CRT beam scans across the film in 2732 discrete vertical steps but the size of the electron beam, or penumbra, in the CRT is larger than the vertical step size resulting in a smoothed image without the pixelated structure of the original image.
%The beam scans across the film in 2732 discrete vertical steps and there is a slight overlap between scan lines ensuring that no digital pixel structure is visible on the film for a smooth analog image. The vertical step between successive horizontal sweeps is intentionally made slightly smaller than the effective beam diameter on the film, so that the “penumbra“ of one line blends with the next. 
Although the process for creating the fork diffraction gratings are the same, differences are observed depending the type of film chosen when investigated under an optical microscope (Leica DM750P microscope, DCM510 CMOS camera). The images of a black and white negative and a colour slide are shown in Figure \ref{fig:gratings}. \\ 
\newpage
\noindent There are two main differences visible at this magnification between the black and white negative and colour slide. In the black and white negative in Figure \ref{fig:gratings}(a), the silver halide grains of the film are clearly visible and the edges of the diffraction grating lines look rough. 
In comparison the colour slide film in Figure \ref{fig:gratings}(b) looks less grainy. 
In the E-6 process for developing positive colour films, the silver halide grains and silver compounds in the film are removed in the fixer stage, therefore the dye clouds are visible instead of the individual grains in the final slide \cite{theys_chemistry_1997,eastman_kodak_company_1_2001}. 
In the dense, black lines of the diffraction grating, the dye clouds blend together giving a smoothed effect and reducing the granularity of the colour film compared to the black and while film \cite{hunt_colour_1977}.
Another feature of the colour slide is that the RGB patterns do not overlap perfectly. 
This is expected to be an unavoidable artifact for the colour slides from the CRT film recorder process and differences in the response of the colour film to red and blue but has not been investigated further to determine the cause. 
Both of the images in Figure \ref{fig:gratings} show that the films are predominately black and blocking the transmission of the beam. The duty cycle of the diffraction lines can be adjusted to produce thinner black lines on the grating in future iterations in order to increase the amount of transmitted light. 
%The limitation of the CRT recording process close to the pixel size of $8.5 \ $\textmu m
%No pixelation is visible from the computer generated design, which sets a limit on the pixel size to $8.5 \ $\textmu m.

\subsection{Experimental realisation of orbital angular momentum beams}
\label{ss:experiment}
\begin{figure}[h!]
    \centering
    \includegraphics[width=0.62\linewidth]{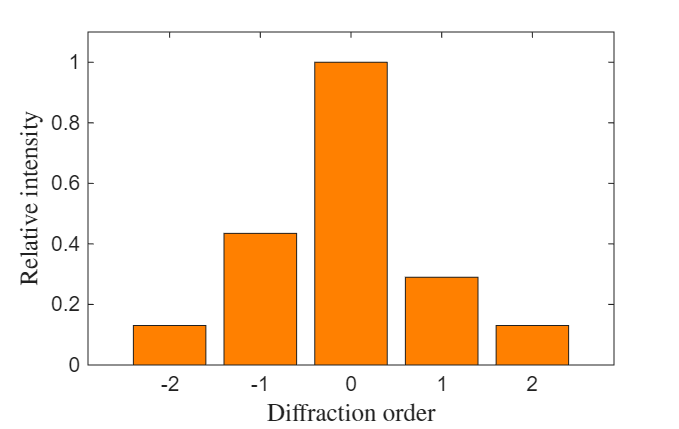}
    \caption{Distribution of intensity in the diffraction orders of the positive colour slide fork diffraction grating, normalised relative to the intensity in the zeroth order of diffraction.}
    \label{fig:diffraction}
\end{figure}

\begin{figure}[h!]
\centering
\begin{subfigure}{0.435\textwidth}
    \centering
    \includegraphics[width=\textwidth]{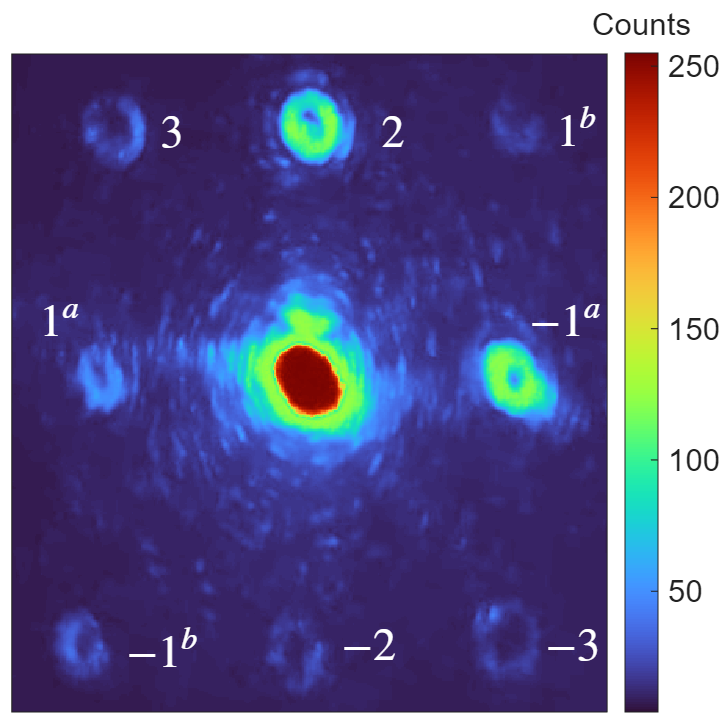}
    \caption{}
    \label{fig:heatmap}
\end{subfigure}
\hfill
\begin{subfigure}{0.545\textwidth}
    \includegraphics[width=\textwidth]{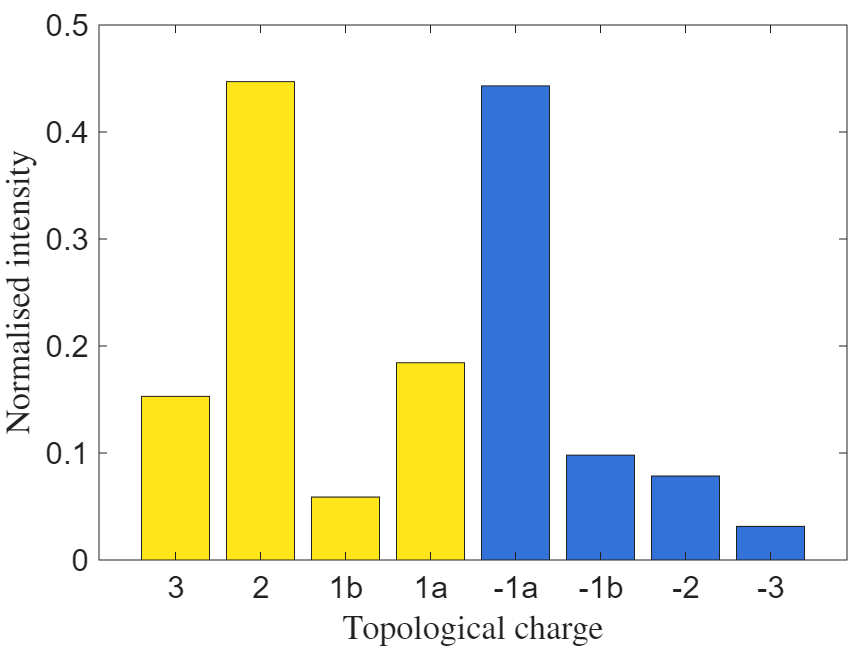}
    \caption{}
    \label{fig:histogram}
\end{subfigure}
\caption{Output and analysis of the beams carrying orbital angular momentum produced by the combined fork diffraction grating on slide film. (a) False colour image of the $3 \times 3$ array produced by the combined fork diffraction grating, with each beam labelled with the charge $\ell$.
%The colour bar is the 8-bit intensity counts of the camera. 
Note that this image is mirrored horizontally compared to Figure \ref{fig:combined_grating} because it is a picture of the screen, not the direct output. (b) The intensity of each topological charge in (a), normalised to the maximum count of 255. }
\label{fig:results_oamarray}
\end{figure}

To test the effectiveness of the film fork diffraction gratings, a simple setup was used, including a laser and a screen or CCD camera. The laser module used was the VLM-650-01 LPA, a Class 3R laser with a maximum power of $3 \ $mW at $650 \ $nm and a spot diameter of $5 \ $mm at $5 \ $m. 
It was powered with a bench top power supply unit, although it can be powered through an Arduino, and provided a laser beam of less than $2 \ $mW incident on the fork diffraction grating.
% Note on the beam shape which caused some problems so lenses and beam expanders were too complicated to use with the setup. 
The laser has a tube casing but no holder, so it was clamped to a kinematic platform (Thorlabs KM100B/M). 
The fork diffraction grating was placed approximately $1 \ $m from the laser, and the screen or camera was placed $1 \ $m after the grating. 
The distances between the components were found to be most important for observing diffraction and OAM beams, rather than the precise alignment of where the laser spot is placed in the region of the fork. Therefore it is expected that 3D printed mounts could work well in the setup because stability in the alignment is not required. 
% These distances depended on the angle of diffraction of the grating and the beam quality of the laser, which became less elliptical and diverged with larger distances. %comment on why a pin hole wasn't used?
\\
With this simple experimental setup, both diffraction and beams carrying OAM were observed. 
Beam shaping lenses, beam expanders, and irises were not used and so the laser beam was elliptical, the beam spot size was 5 mm at a distance of 5 m, and the beam divergence half-angle was 0.5 mrad. However, OAM beams were still observed, showing that the film diffraction gratings are robust to the absence of beam shaping lenses. 
The distribution of laser light in the first two orders of diffraction from a fork diffraction grating with period spacing of $50 \ $\textmu m and input laser power of $\sim 1.5 \ $mW is shown in Figure \ref{fig:diffraction}.\\
\\ 
The relative intensity of the first orders of diffraction were not equal, showing some asymmetry in the setup due to alignment of the laser on the grating or the beam shape.
Both the slide and negative film fork diffraction gratings were tested, and there were no clear differences observed between the diffraction gratings on each type of film. \\
Further, the OAM beams produced by the photographic film fork diffraction gratings were analysed.
Figure \ref{fig:results_oamarray} shows the $3 \times 3$ array created by the combined diffraction grating and a histogram of counts of each OAM beam, normalised to the maximum 8-bit count value. 
The Gaussian mode in the centre is the brightest and saturates the camera, but all of the eight beams carrying OAM are visible on the screen when looking through a camera.
\\
The intensities are inhomogeneous and very asymmetric, especially for $\ell = 2$ and $\ell = -2$ in Figure \ref{fig:results_oamarray}. Further measurements and analysis of the combined grating are presented in Appendix \ref{s:supfig}. 
Future iterations of the combined fork diffraction grating will explore ways to produce a more homogeneous distribution and higher intensity in all of the OAM beams. The combined diffraction grating may also benefit from adding beam shaping lenses and iris to create a symmetric TEM$_{00}$ input laser beam, at the expense of increasing the complexity of the setup and the cost.

\section{Classroom demonstrations} % Demonstrations for education in school and undergraduate programmes (school instead of college or a-level, more univeral word)
Given the results of OAM beams produced by the photographic film diffraction gratings, the experimental setup as described above is suitable as a research-based learning demonstration for OAM. 
Adjustments to the experimental setup are required for the safe use of equipment by students, and these are discussed in this section.

\subsection{Demonstration in the classroom or lecture theatre}
By mounting the fork diffraction grating and laser on a breadboard, or mounting directly to the table or desk, the OAM beams can be shown to students as a research-based learning demonstration. 
Research shows that involving students in experiments can positively influence interest in STEM topics, more so compared to answer-led approaches \cite{toli_enhancing_2021,munoz-losa_impact_2025} and increase performance compared to traditional lecturing \cite{freeman_active_2014}.
One way to involve students in the demonstration is to ask for predictions of the outcome of the experiment.
Therefore, key concepts and theories should be covered before presenting the demonstration, such as diffraction and OAM, for effective teaching \cite{miller_role_2013}. 
 \\
To make the experiment suitable for the classroom, a Class 2 laser pointer at $650 \ $nm with a maximum power of $1 \ $mW is required.
With low laser power and in a well lit classroom, it may be difficult to see the laser spot on a screen, and a camera may be required. 
%Using a camera is advantageous for certain situations, since the camera output can be projected on the board for large lectures, or for situations where it is difficult to move around, recording on a computer for use in asynchronous videos.
A camera is also advantageous for certain situations where it is difficult or not possible for students to move around to observe the demonstration. For example, the camera output can be projected on the board for large lectures, or the output can be recorded on a computer for use in asynchronous videos.
A phone camera is found to work well for observing the laser spots at 650 nm.
% A selection of diffraction gratings, fork diffraction gratings of charges up to $\ell=3$, and combined diffraction gratings are provided as jpg images at 
%bristol repo?
%on Github at \href{https://github.com/OAM-MSC-OQT-MATLAB/Orbital-angular-momentum-mask-and-simulations}{github.com/OAM-MSC-OQT-MATLAB}.

\subsection{Laboratory activity}
The use of low-cost film fork diffraction gratings and Class 2 laser pointers presents an interesting and cost-effective laboratory activity for groups of students at school and undergraduate level.
For introducing quantum mechanics concepts and OAM, the setup demonstrated in Section \ref{ss:experiment} is effective as an investigative science learning environment (ISLE) activity \cite{etkina_investigative_2020}.
Using this method for teaching quantum mechanics concepts is novel in a topic where analogies, gamification \cite{chiofalo_games_2024}, and virtual learning environments \cite{pedersen_virtual_2016} are usually used. 
In previous studies employing ISLE lab activities, students were found to have a better understanding of concepts in order to meet learning outcomes, and a perception that they understood the concepts better \cite{wilson_teaching_2020}. 
Students also gained contextualised laboratory skills, which are important skills for undergraduates on physics and engineering courses to have when entering optics and quantum mechanical fields \cite{aiello_achieving_2021}.
The ISLE method works with instructors running the activity with more open-ended questions and few instructions, and the students leading the experiments, making their own hypotheses and observations that support or debunk their theories.
Another advantage of the ISLE method is the experiment is the flexibility offered to cater for various groups of students \cite{etkina_investigative_2020}. 
The activity can be introduced with little to no maths, making it more accessible to students from any background \cite{wilson_teaching_2020} and interdisciplinary subject areas, and maths can be added as desired for physics and optics-focused students. \\
For school students, a common diffraction activity can be done prior to introducing the production of OAM beams, such as determining the wavelength of the laser pointer with a diffraction grating of known period. Given the number of lines per mm that can be achieved with the film diffraction gratings, the experiment can be performed on the scale of a table rather than across a room, which is a limitation for printed diffraction gratings \cite{van_hook_inquiry_2007}. 
An example laboratory activity worksheet aimed at school students is included in the supplementary files.

\section{Conclusions}
This work has demonstrated that photographic film fork diffraction gratings can induce beams carrying OAM from an incident Gaussian beam at $650 \ $nm, and it is possible to produce complex designs on film such as combined fork diffraction gratings.
Artifacts in the black and white negative film and colour slides, such as the granularity of the film and misaligned RGB colours respectively, did not prevent the production of light carrying OAM. 
Improvements to the beam shape with cylindrical lenses to make the Gaussian mode more spherical are expected to improve the efficiency of the OAM beams produced, however, the gratings were robust to an asymmetric input beam and misalignments.
The low-cost and commercial availability of fork diffraction gratings on photographic film make them suitable as a demonstration in a lecture or classroom, and as an ISLE laboratory activity. \\
The initial results shown in this paper suggest that improvements can be made with more iterations of the fork diffraction gratings, including variations in the period spacing and duty cycles of the grating.
The limit of the resolution and performance of the film diffraction gratings with computer generated holograms at the single pixel level are particularity interesting, in order to reduce the distances required for observing significant separation of the diffraction orders.
%limitations of the size of gratings as well, down to single pixel level
In the future, the tools will be implemented as demonstrations in lectures and as an activity for school visits in the School of Electrical, Electronic and Mechanical Engineering and the School of Physics at the University of Bristol.

\section*{Acknowledgements}
The authors are grateful to Chris Brown and Mariana Reyes for facilitating the use of the microscope in the Materials lab in the School of Physics, University of Bristol. 
%include contributions list? otherwise all considered equal contribution

\section*{Funding}
This work was supported by the Engineering and Physical Sciences Research Council (EPSRC) New Investigator Award grant number EP/X029360/1, the Royal Academy of Engineering (RAEng) and the Department of Science, Innovation and Technology (DSIT) Industry Fellowship (Industry to Academia) IF2425-19-IA-110, and Innovate UK Quantum Missions Pilot: Quantum Computing and Quantum Networks, IUK: 10149275.

% \section*{Author contributions} % to be removed for the submitted article, seperate credit form is submitted
% N Bierrum: Writing- original draft, conceptualisation, investigation, visualisation, supervision; L Chen: Writing- original draft, visualisation, investigation; A Kudaloor, LK Wan, S Yang, Y Hou, X Dong, M Tuli: investigation; R Taylor, P Androvitsaneas, C Weidner: conceptualisation; E Harbord: Writing- original draft, conceptualisation, supervision.
% All authors: Writing - review \& editing.

\printbibliography

@article{allen_orbital_1992,
	title = {Orbital angular momentum of light and the transformation of {Laguerre}-{Gaussian} laser modes},
	volume = {45},
	doi = {10.1103/PhysRevA.45.8185},
	pages = {8185--8189},
	number = {11},
	journal = {Physical Review A},
	shortjournal = {Phys. Rev. A},
	author = {Allen, L. and Beijersbergen, M. W. and Spreeuw, R. J. C. and Woerdman, J. P.},
	year = {1992},
}

@article{yao_orbital_2011,
	title = {Orbital angular momentum: origins, behavior and applications},
	volume = {3},
	rights = {https://doi.org/10.1364/{OA}\_License\_v1\#{VOR}},
	issn = {1943-8206},
	doi = {10.1364/AOP.3.000161},
	shorttitle = {Orbital angular momentum},
	pages = {161},
	number = {2},
	journal = {Advances in Optics and Photonics},
	shortjournal = {Adv. Opt. Photon.},
	author = {Yao, Alison M. and Padgett, Miles J.},
	year = {2011},
}

@article{zheng_measuring_2017,
	title = {{Measuring Orbital Angular Momentum ({OAM}) States of Vortex Beams with Annular Gratings}},
	volume = {7},
	rights = {2017 The Author(s)},
	issn = {2045-2322},
	doi = {10.1038/srep40781},
	pages = {40781},
	number = {1},
	journal = {Scientific Reports},
	shortjournal = {Sci Rep},
	author = {Zheng, Shuang and Wang, Jian},
	year = {2017},
}

@article{stoyanov_far_2015,
	title = {Far field diffraction of an optical vortex beam by a fork-shaped grating},
	volume = {350},
	issn = {00304018},
	doi = {10.1016/j.optcom.2015.04.020},
	pages = {301--308},
	journal = {Optics Communications},
	author = {Stoyanov, Lyubomir and Topuzoski, Suzana and Stefanov, Ivan and Janicijevic, Ljiljana and Dreischuh, Alexander},
	year = {2015},
}

@article{greinert_future_2023,
	title = {{Future quantum workforce: Competences, requirements, and forecasts}},
	volume = {19},
	issn = {2469-9896},
	doi = {10.1103/PhysRevPhysEducRes.19.010137},
	shorttitle = {Future quantum workforce},
	pages = {010137},
	number = {1},
	journaltitle = {Physical Review Physics Education Research},
	shortjournal = {Phys. Rev. Phys. Educ. Res.},
	author = {Greinert, Franziska and Müller, Rainer and Bitzenbauer, Philipp and Ubben, Malte S. and Weber, Kim-Alessandro},
	year = {2023},
}

@article{larocque_generalized_2017,
	title = {Generalized optical angular momentum sorter and its application to high-dimensional quantum cryptography},
	volume = {25},
	rights = {© 2017 Optical Society of America},
	issn = {1094-4087},
	doi = {10.1364/OE.25.019832},
	pages = {19832--19843},
	number = {17},
	journaltitle = {Optics Express},
	shortjournal = {Opt. Express, {OE}},
	publisher = {Optica Publishing Group},
	author = {Larocque, Hugo and Gagnon-Bischoff, Jérémie and Mortimer, Dominic and Zhang, Yingwen and Bouchard, Frédéric and Upham, Jeremy and Grillo, Vincenzo and Boyd, Robert W. and Karimi, Ebrahim},
	year = {2017},
}

@article{freeman_active_2014,
	title = {Active learning increases student performance in science, engineering, and mathematics},
	volume = {111},
	issn = {0027-8424, 1091-6490},
	doi = {10.1073/pnas.1319030111},
	pages = {8410--8415},
	number = {23},
	journaltitle = {Proceedings of the National Academy of Sciences},
	shortjournal = {Proc. Natl. Acad. Sci. U.S.A.},
	author = {Freeman, Scott and Eddy, Sarah L. and {McDonough}, Miles and Smith, Michelle K. and Okoroafor, Nnadozie and Jordt, Hannah and Wenderoth, Mary Pat},
	year = {2014},
}

@incollection{eastman_kodak_company_1_2001,
	edition = {6},
	title = {Processing solutions and their effects},
	series = {Z-119},
	booktitle = {Using {KODAK} chemicals, Process E-6},
	publisher = {Kodak},
	%author = {Eastman Kodak Company},
	date = {2001},
}

@incollection{etkina_investigative_2020,
	title = {{Investigative Science Learning Environment: Learn Physics by Practicing Science}},
	isbn = {978-3-030-33600-4},
	doi = {10.1007/978-3-030-33600-4_23},
	shorttitle = {Investigative Science Learning Environment},
	pages = {359--383},
	booktitle = {Active Learning in College Science: The Case for Evidence-Based Practice},
	publisher = {Springer International Publishing},
	author = {Etkina, Eugenia and Brookes, David T. and Planinsic, Gorazd},
	editor = {Mintzes, Joel J. and Walter, Emily M.},
	year = {2020},
}

@article{mirhosseini_high-dimensional_2015,
	title = {High-dimensional quantum cryptography with twisted light},
	volume = {17},
	issn = {1367-2630},
	doi = {10.1088/1367-2630/17/3/033033},
	pages = {033033},
	number = {3},
	journaltitle = {New Journal of Physics},
	publisher = {{IOP} Publishing},
	author = {Mirhosseini, Mohammad and Magaña-Loaiza, Omar S and O’Sullivan, Malcolm N and Rodenburg, Brandon and Malik, Mehul and Lavery, Martin P J and Padgett, Miles J and Gauthier, Daniel J and Boyd, Robert W},
	year = {2015},
}

@article{chen_orbital_2020,
	title = {{Orbital Angular Momentum Waves: Generation, Detection, and Emerging Applications}},
	volume = {22},
	issn = {1553-877X},
	doi = {10.1109/COMST.2019.2952453},
	shorttitle = {Orbital Angular Momentum Waves},
	pages = {840--868},
	number = {2},
	journaltitle = {{IEEE} Communications Surveys \& Tutorials},
	author = {Chen, Rui and Zhou, Hong and Moretti, Marco and Wang, Xiaodong and Li, Jiandong},
	year = {2020},
}

@article{willner_optical_2015,
	title = {Optical communications using orbital angular momentum beams},
	volume = {7},
	rights = {https://doi.org/10.1364/{OA}\_License\_v1\#{VOR}},
	issn = {1943-8206},
	doi = {10.1364/AOP.7.000066},
	pages = {66},
	number = {1},
	journal = {Advances in Optics and Photonics},
	shortjournal = {Adv. Opt. Photon.},
	author = {Willner, A. E. and Huang, H. and Yan, Y. and Ren, Y. and Ahmed, N. and Xie, G. and Bao, C. and Li, L. and Cao, Y. and Zhao, Z. and Wang, J. and Lavery, M. P. J. and Tur, M. and Ramachandran, S. and Molisch, A. F. and Ashrafi, N. and Ashrafi, S.},
	year = {2015},
}

@article{gibson_free-space_2004,
	title = {Free-space information transfer using light beams carrying orbital angular momentum},
	volume = {12},
	rights = {© 2004 Optical Society of America},
	issn = {1094-4087},
	doi = {10.1364/OPEX.12.005448},
	pages = {5448--5456},
	number = {22},
	journal = {Optics Express},
	shortjournal = {Opt. Express, {OE}},
	author = {Gibson, Graham and Courtial, Johannes and Padgett, Miles J. and Vasnetsov, Mikhail and Pas’ko, Valeriy and Barnett, Stephen M. and Franke-Arnold, Sonja},
	year = {2004},
}

@online{digital_2025,
	title = {Digital {Slides}},
	url = {https://www.digitalslides.co.uk/product-category/35mm/},
	type = {Online shop},
    date = {2025}
}

@article{seifollahi_preparing_2025,
	title = {Preparing students for the quantum information revolution: interdisciplinary teaching, curriculum development, and advising in quantum information science and engineering},
	volume = {46},
	issn = {0143-0807, 1361-6404},
	doi = {10.1088/1361-6404/ae0200},
	shorttitle = {Preparing students for the quantum information revolution},
	pages = {055709},
	number = {5},
	journal = {European Journal of Physics},
	shortjournal = {Eur. J. Phys.},
	author = {Seifollahi, Fargol and Singh, Chandralekha},
	year = {2025},
}

@article{miller_role_2013,
	title = {Role of physics lecture demonstrations in conceptual learning},
	volume = {9},
	rights = {http://creativecommons.org/licenses/by/3.0/},
	issn = {1554-9178},
	doi = {10.1103/PhysRevSTPER.9.020113},
	pages = {020113},
	number = {2},
	journaltitle = {Physical Review Special Topics - Physics Education Research},
	shortjournal = {Phys. Rev. {ST} Phys. Educ. Res.},
	author = {Miller, Kelly and Lasry, Nathaniel and Chu, Kelvin and Mazur, Eric},
	year = {2013},
}

@article{yang_review_2023,
	title = {A review of liquid crystal spatial light modulators: devices and applications},
	volume = {2},
	rights = {http://creativecommons.org/licenses/by/3.0/},
	issn = {2097-0382},
	doi = {10.29026/oes.2023.230026},
	shorttitle = {A review of liquid crystal spatial light modulators},
	pages = {230026--29},
	number = {8},
	journaltitle = {Opto-Electronic Science},
	shortjournal = {Opto-Electron Sci},
	author = {Yang, Yiqian and Forbes, Andrew and Cao, Liangcai},
	year = {2023},
}

@article{chiofalo_games_2024,
	title = {{Games for Quantum Physics Education}},
	volume = {2727},
	issn = {1742-6596},
	doi = {10.1088/1742-6596/2727/1/012010},
	pages = {012010},
	number = {1},
	journaltitle = {Journal of Physics: Conference Series},
	shortjournal = {J. Phys.: Conf. Ser.},
	publisher = {{IOP} Publishing},
	author = {Chiofalo, Maria Luisa and Foti, Caterina and Lazzeroni, Cristina and Maniscalco, Sabrina and Seskir, Zeki C. and Sherson, Jacob and Weidner, Carrie Ann and Michelini, Marisa},
	year = {2024-03},
}

@article{pedersen_virtual_2016,
	title = {{Virtual Learning Environment for Interactive Engagement with Advanced Quantum Mechanics}},
	volume = {12},
	rights = {http://creativecommons.org/licenses/by/3.0/},
	issn = {2469-9896},
	doi = {10.1103/PhysRevPhysEducRes.12.013102},
	pages = {013102},
	number = {1},
	journaltitle = {Physical Review Physics Education Research},
	shortjournal = {Phys. Rev. Phys. Educ. Res.},
	author = {Pedersen, Mads Kock and Skyum, Birk and Heck, Robert and Müller, Romain and Bason, Mark and Lieberoth, Andreas and Sherson, Jacob F.},
	year = {2016},
}

@article{vaziri_simple_2024,
	title = {A simple method to prepare and characterize optical fork-shaped diffraction gratings for generation of orbital angular momentum beams},
	issn = {0974-6900},
	doi = {10.1007/s12596-024-02154-9},
	journal = {Journal of Optics},
	shortjournal = {J Opt},
	author = {Vaziri, Mohammad Reza Rashidian and Hosseini, Abolfazl and Hatam, Ebrahim Gholami and Sorodi, Reza Azmoodeh},
	year = {2024},
}

@article{malik_influence_2012,
	title = {Influence of atmospheric turbulence on optical communications using orbital angular momentum for encoding},
	volume = {20},
	rights = {https://doi.org/10.1364/{OA}\_License\_v1\#{VOR}-{OA}},
	issn = {1094-4087},
	doi = {10.1364/OE.20.013195},
	pages = {13195},
	number = {12},
	journaltitle = {Optics Express},
	shortjournal = {Opt. Express},
	author = {Malik, Mehul and O’Sullivan, Malcolm and Rodenburg, Brandon and Mirhosseini, Mohammad and Leach, Jonathan and Lavery, Martin P. J. and Padgett, Miles J. and Boyd, Robert W.},
	year = {2012},
}

@article{dai_measuring_2015,
	title = {Measuring {OAM} states of light beams with gradually-changing-period gratings},
	volume = {40},
	rights = {https://doi.org/10.1364/{OA}\_License\_v1\#{VOR}},
	issn = {0146-9592, 1539-4794},
	doi = {10.1364/OL.40.000562},
	pages = {562},
	number = {4},
	journal = {Optics Letters},
	shortjournal = {Opt. Lett.},
	author = {Dai, Kunjian and Gao, Chunqing and Zhong, Lei and Na, Quanxin and Wang, Qing},
	year = {2015},
}

@article{wilson_teaching_2020,
	title = {Teaching physics concepts without much mathematics: ensuring physics is available to students of all backgrounds},
	volume = {25},
	issn = {2205-4952, 1325-4340},
	doi = {10.1080/22054952.2020.1776027},
	shorttitle = {Teaching physics concepts without much mathematics},
	pages = {39--54},
	number = {1},
	journal = {Australasian Journal of Engineering Education},
	shortjournal = {Australasian Journal of Engineering Education},
	author = {Wilson, Marcus T. and Seshadri, Sinduja and Streeter, Lee V. and Scott, Jonathan B.},
	year = {2020},
}

@inproceedings{ii_increasing_2004,
	title = {Increasing the data density of free-space optical communications using orbital angular momentum},
	volume = {5550},
	doi = {10.1117/12.557176},
	eventtitle = {Free-Space Laser Communications {IV}},
	pages = {367--373},
	booktitle = {Free-Space Laser Communications {IV}},
	publisher = {{SPIE}},
	author = {Ii, Graham Gibson and Courtial, Johannes and Vasnetsov, Mikhail and Barnett, Steve and Franke-Arnold, Sonja and Padgett, Miles},
	year = {2004},
}

@article{lee_fabrication_2010,
	title = {Fabrication of slits for {Young’s} experiment using graphic arts films},
	volume = {78},
	issn = {0002-9505},
	doi = {10.1119/1.3230035},
	pages = {71--74},
	number = {1},
	journal = {American Journal of Physics},
	shortjournal = {Am. J. Phys.},
	author = {Lee, Changsug and Shin, Kwangmoon and Lee, Sungmuk and Lee, Jaebong},
	year = {2010},
}

@article{van_hook_inquiry_2007,
	title = {{Inquiry with Laser Printer Diffraction Gratings}},
	volume = {45},
	issn = {0031-921X},
	doi = {10.1119/1.2768688},
	pages = {340--343},
	number = {6},
	journal = {The Physics Teacher},
	shortjournal = {Phys. Teach.},
	author = {Van Hook, Stephen J.},
	year = {2007},
}

@article{theys_chemistry_1997,
	title = {{Chemistry and Processes of Color Photography}},
	volume = {97},
	issn = {0009-2665, 1520-6890},
	doi = {10.1021/cr941191p},
	pages = {83--132},
	number = {1},
	journaltitle = {Chemical Reviews},
	shortjournal = {Chem. Rev.},
	author = {Theys, Ronald D. and Sosnovsky, George},
	year = {1997},
}

@article{hunt_colour_1977,
	title = {Colour reproduction by photography},
	volume = {40},
	issn = {0034-4885},
	doi = {10.1088/0034-4885/40/9/003},
	pages = {1071},
	number = {9},
	journaltitle = {Reports on Progress in Physics},
	shortjournal = {Rep. Prog. Phys.},
	author = {Hunt, R. W. G.},
	year = {1977},
}

@article{toli_enhancing_2021,
	title = {{Enhancing Student Interest to Promote Learning in Science: The Case of the Concept of Energy}},
	volume = {11},
	rights = {http://creativecommons.org/licenses/by/3.0/},
	issn = {2227-7102},
	doi = {10.3390/educsci11050220},
	shorttitle = {Enhancing Student Interest to Promote Learning in Science},
	pages = {220},
	number = {5},
	journaltitle = {Education Sciences},
	publisher = {Multidisciplinary Digital Publishing Institute},
	author = {Toli, Georgia and Kallery, Maria},
	year = {2021},
}

@article{munoz-losa_impact_2025,
	title = {{Impact of Interactive Science Workshops Participation on Primary School Children's Emotions and Attitudes Towards Science}},
	volume = {23},
	issn = {1573-1774},
	doi = {10.1007/s10763-024-10539-2},
	pages = {2689--2706},
	number = {7},
	journaltitle = {International Journal of Science and Mathematics Education},
	shortjournal = {Int J of Sci and Math Educ},
	author = {Muñoz-Losa, Aurora and Corbacho-Cuello, Isaac},
	year = {2025},
}

@article{fox_preparing_2020,
	title = {{Preparing for the quantum revolution: What is the role of higher education?}},
	volume = {16},
	issn = {2469-9896},
	doi = {10.1103/PhysRevPhysEducRes.16.020131},
	shorttitle = {Preparing for the quantum revolution},
	pages = {020131},
	number = {2},
	journaltitle = {Physical Review Physics Education Research},
	shortjournal = {Phys. Rev. Phys. Educ. Res.},
	author = {Fox, Michael F.J. and Zwickl, Benjamin M. and Lewandowski, H.J.},
	year = {2020},
}

@article{velentzas_teaching_2014,
	title = {{Teaching Diffraction of Light and Electrons: Classroom Analogies to Classic Experiments}},
	volume = {52},
	issn = {0031-921X, 1943-4928},
	doi = {10.1119/1.4897589},
	shorttitle = {Teaching Diffraction of Light and Electrons},
	pages = {493--496},
	number = {8},
	journaltitle = {The Physics Teacher},
    shortjournal = {Phys. Teach.},
	author = {Velentzas, Athanasios},
	year = {2014},
}

@article{aiello_achieving_2021,
	title = {Achieving a quantum smart workforce},
	volume = {6},
	issn = {2058-9565},
	doi = {10.1088/2058-9565/abfa64},
	pages = {030501},
	number = {3},
	journaltitle = {Quantum Science and Technology},
	shortjournal = {Quantum Sci. Technol.},
	publisher = {{IOP} Publishing},
	author = {Aiello, Clarice D and Awschalom, D D and Bernien, Hannes and Brower, Tina and Brown, Kenneth R and Brun, Todd A and Caram, Justin R and Chitambar, Eric and Di Felice, Rosa and Edmonds, Karina Montilla and Fox, Michael F J and Haas, Stephan and Holleitner, Alexander W and Hudson, Eric R and Hunt, Jeffrey H and Joynt, Robert and Koziol, Scott and Larsen, M and Lewandowski, H J and {McClure}, Doug T and Palsberg, Jens and Passante, Gina and Pudenz, Kristen L and Richardson, Christopher J K and Rosenberg, Jessica L and Ross, R S and Saffman, Mark and Singh, M and Steuerman, David W and Stark, Chad and Thijssen, Jos and Vamivakas, A Nick and Whitfield, James D and Zwickl, Benjamin M},
	year = {2021},
}

@article{beijersbergen_helical-wavefront_1994,
	title = {Helical-wavefront laser beams produced with a spiral phaseplate},
	volume = {112},
	issn = {0030-4018},
	doi = {10.1016/0030-4018(94)90638-6},
	pages = {321--327},
	number = {5},
	journal= {Optics Communications},
	shortjournal = {Optics Communications},
	author = {Beijersbergen, M. W. and Coerwinkel, R. P. C. and Kristensen, M. and Woerdman, J. P.},
	year = {1994},
}

@article{heckenberg_generation_1992,
	title = {Generation of optical phase singularities by computer-generated holograms},
	volume = {17},
	rights = {© 1992 Optical Society of America},
	issn = {1539-4794},
	doi = {10.1364/OL.17.000221},
	pages = {221--223},
	number = {3},
	journal = {Optics Letters},
	shortjournal = {Opt. Lett., {OL}},
	author = {Heckenberg, N. R. and {McDuff}, R. and Smith, C. P. and White, A. G.},
	year = {1992},
}

@article{oemrawsingh_production_2004,
	title = {Production and characterization of spiral phase plates for optical wavelengths},
	volume = {43},
	rights = {https://doi.org/10.1364/{OA}\_License\_v1\#{VOR}},
	issn = {0003-6935, 1539-4522},
	doi = {10.1364/AO.43.000688},
	pages = {688},
	number = {3},
	journal = {Applied Optics},
	shortjournal = {Appl. Opt.},
	author = {Oemrawsingh, S. S. R. and Van Houwelingen, J. A. W. and Eliel, E. R. and Woerdman, J. P. and Verstegen, E. J. K. and Kloosterboer, J. G. and ’T Hooft, G. W.},
	year = {2004},
}
\newpage
\appendix
\section{Laguerre-Gaussian modes}
\label{s:LGmode}
The amplitude distribution $A_p^l(r,z)$ of the Laguerre-Gaussian modes, $LG_{p}^\ell (r, \phi, z) = \\ A_{p}^\ell (r,z) \exp \left( i \ell \phi \right)$, is
\begin{align*}
    A_p^l(r,z) &= \sqrt{\frac{2p!}{\pi \left(p+\left| \ell \right| \right)!}} \frac{1}{w(z)} \left( \frac{r \sqrt{2}}{w(z)} \right) ^{\left| \ell \right|} \exp \left[ \frac{-r^{2}}{w^2(z)} \right] L_p^{\left| \ell \right|} \left( \frac{2r^{2}}{w^2(z)} \right) \exp \left[ \frac{ik_0r^{2}z}{2(z^2 + z_R^2)} \right] \nonumber \\ 
    & \exp \left[ -i(2p + \left| \ell \right| + 1) \tan^{-1} \left( \frac{z}{z_R} \right) \right],
\end{align*}
where the $1/e$ radius of the Gaussian component %is this clear enough?
is $w(z) = w(0) \left((z^2 + z_R^2)/z_R^2\right)^{1/2}$, $w(0)$ is the beam waist, $z_R$ is the Rayleigh range, $k_0 = 2 \pi/ \lambda$, $\lambda$ is the wavelength of the light, $ L_p^{\left| \ell \right|}$ is the associated Laguerre polynomial, and the Gouy phase for the $LG$ mode is $(2p + \left| \ell \right| + 1) \tan^{-1} \left( \sfrac{z}{z_R} \right)$. The Laguerre polynomials are given by 
\begin{equation*}
    L_p^{\left| \ell \right|} \left( x \right) = (-1)^{\left| \ell \right|}  \frac{\textrm{d}^{\left| l \right|}}{\textrm{d} x^{\left| \ell \right|}} L_{p+\left| \ell \right|} \left( x \right), 
\end{equation*}
where $\ell$ is azimuthal wavefront number, or topological charge, $\ell \in \mathbb{Z}$, and $p$ is the number of radial nodes in the intensity distribution.
\section{Additional analysis of the combined diffraction grating}
\label{s:supfig}
%check what description is needed
\begin{figure}[h!]
    \centering
    \includegraphics[width=0.5\linewidth]{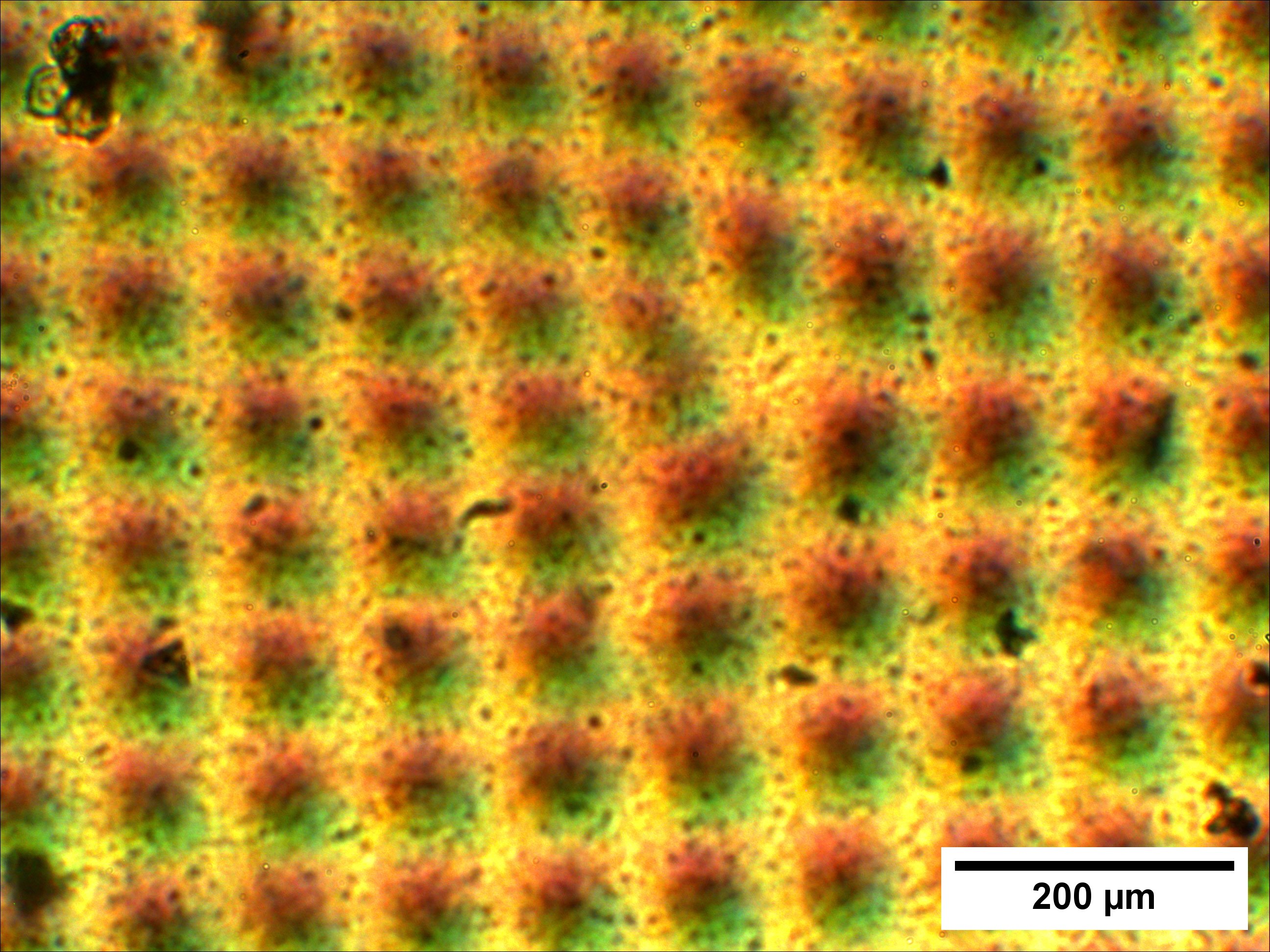} 
    \caption{Microscope image of the superimposed fork diffraction grating.}
    \label{fig:combined_grating_microscope}
\end{figure}

\begin{figure}    
  \begin{subfigure}[b]{0.48\linewidth}
    \centering
    \includegraphics[height=0.2\textheight]{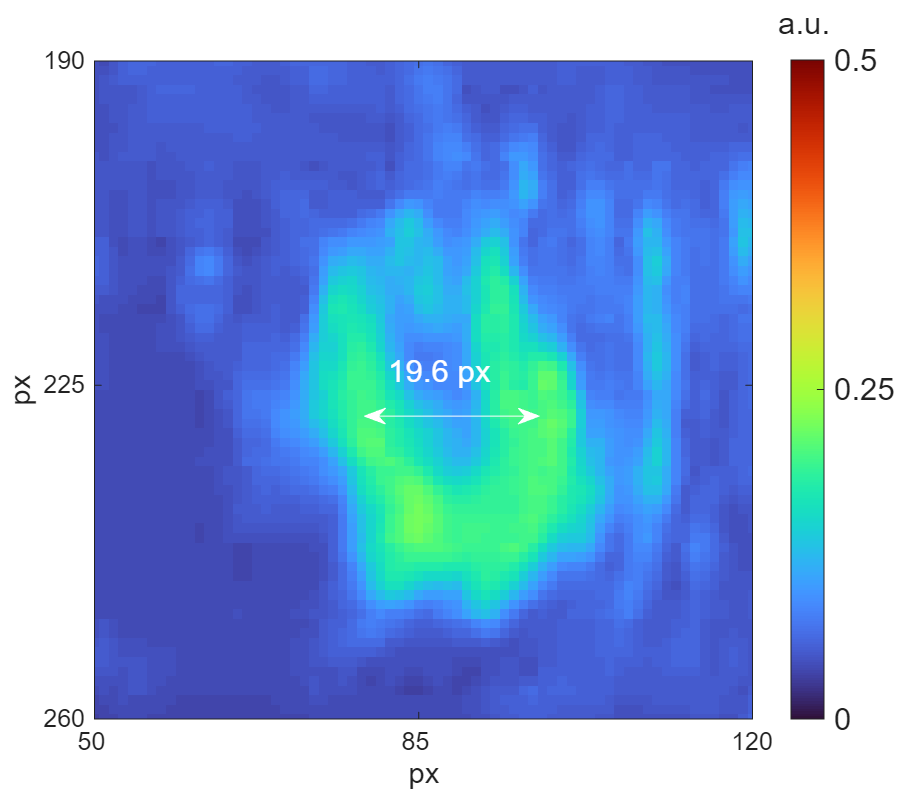} 
    \caption{} 
    \label{fig:l1a} 
  \end{subfigure} 
  \hspace{\fill}  %% maximize space between adjacent subfigures
  \begin{subfigure}[b]{0.48\linewidth}
    \centering
    \includegraphics[height=0.2\textheight]{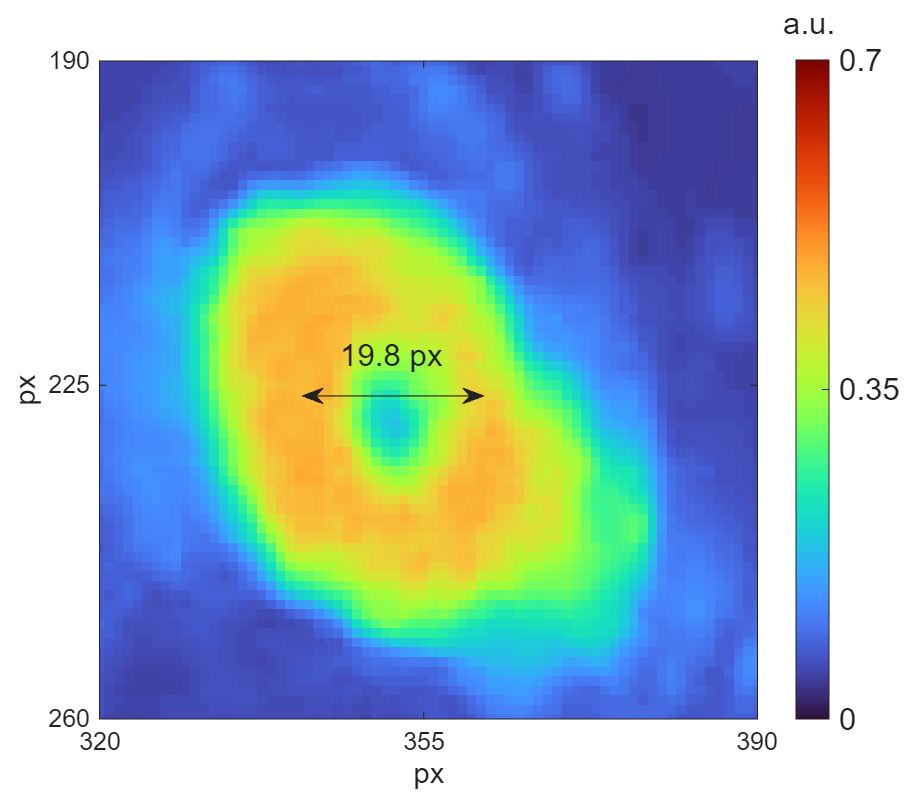} 
    \caption{} 
    \label{fig:lm1a} 
  \end{subfigure} 

  \vspace{1ex}  %% extra vertical space
  \begin{subfigure}[b]{0.48\linewidth}
    \centering
    \includegraphics[height=0.2\textheight]{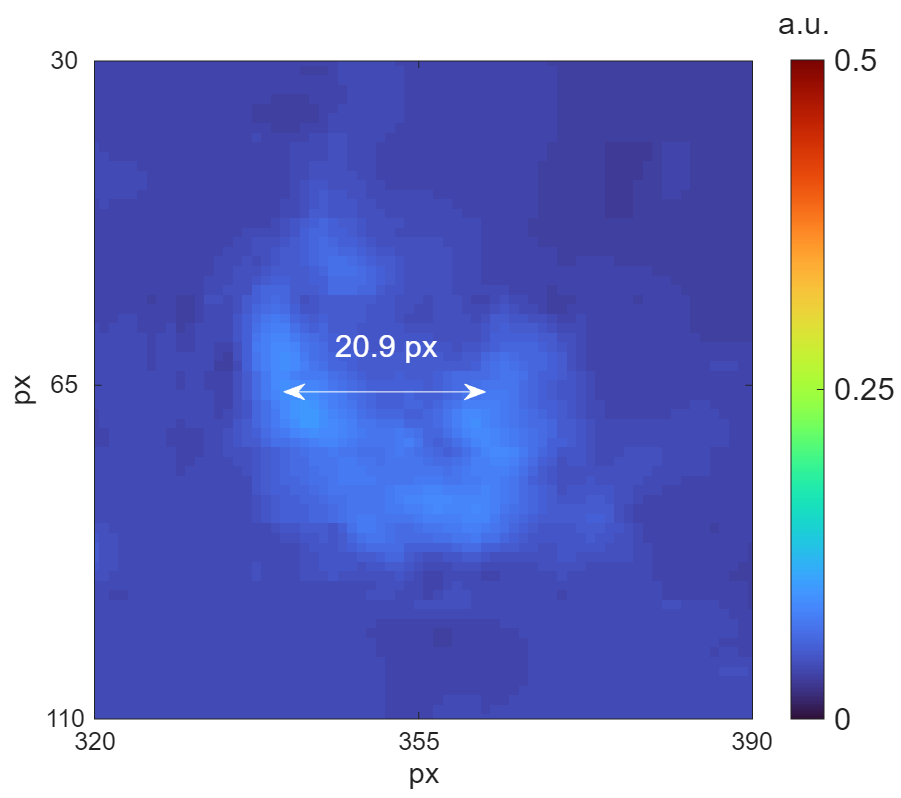} 
    \caption{} 
    \label{fig:l1b} 
  \end{subfigure} 
  \hspace{\fill}
  \begin{subfigure}[b]{0.48\linewidth}
    \centering
    \includegraphics[height=0.2\textheight]{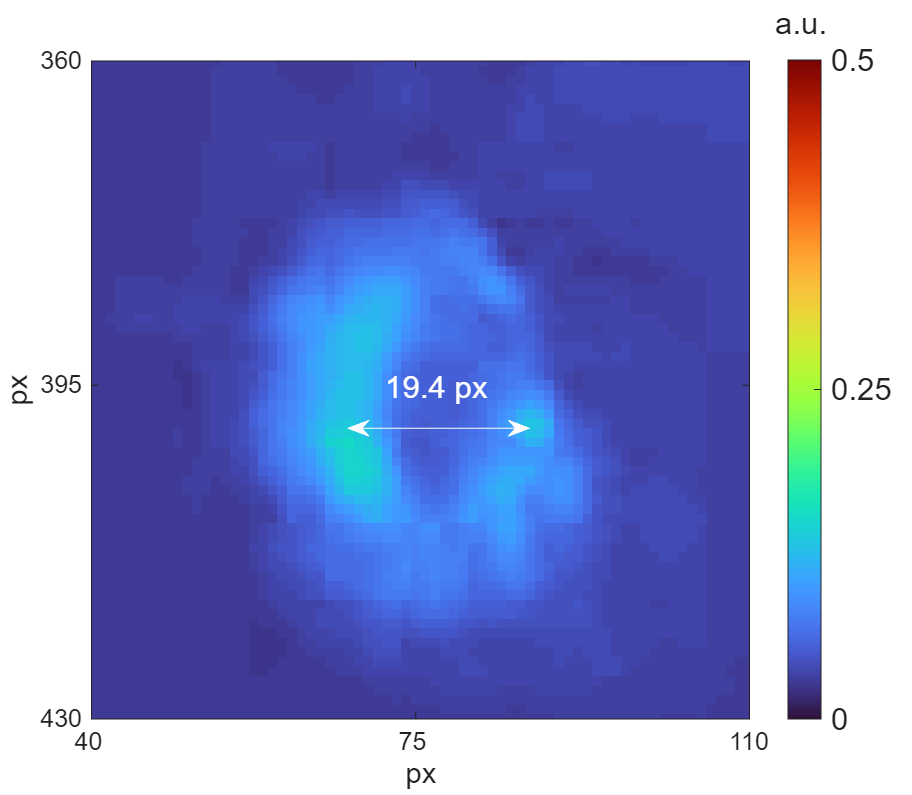} 
    \caption{} 
    \label{fig:lm1b} 
  \end{subfigure} 

    \vspace{1ex}
  \begin{subfigure}[b]{0.48\linewidth}
    \centering
    \includegraphics[height=0.2\textheight]{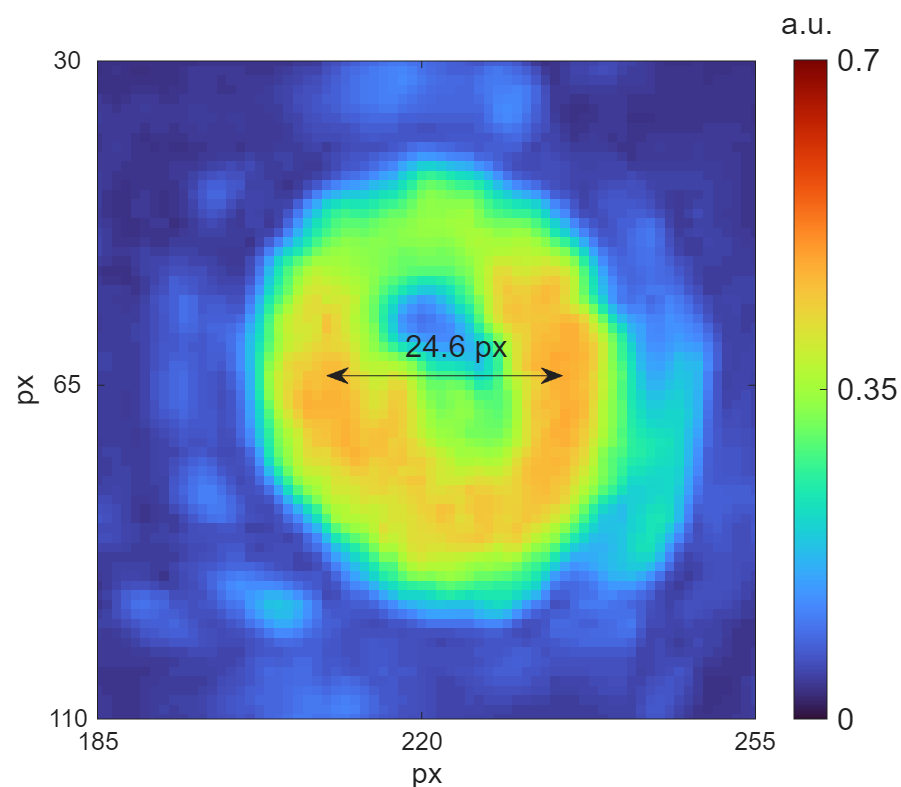} 
    \caption{} 
    \label{fig:l2} 
  \end{subfigure}
  \hspace{\fill}
  \begin{subfigure}[b]{0.48\linewidth}
    \centering
    \includegraphics[height=0.2\textheight]{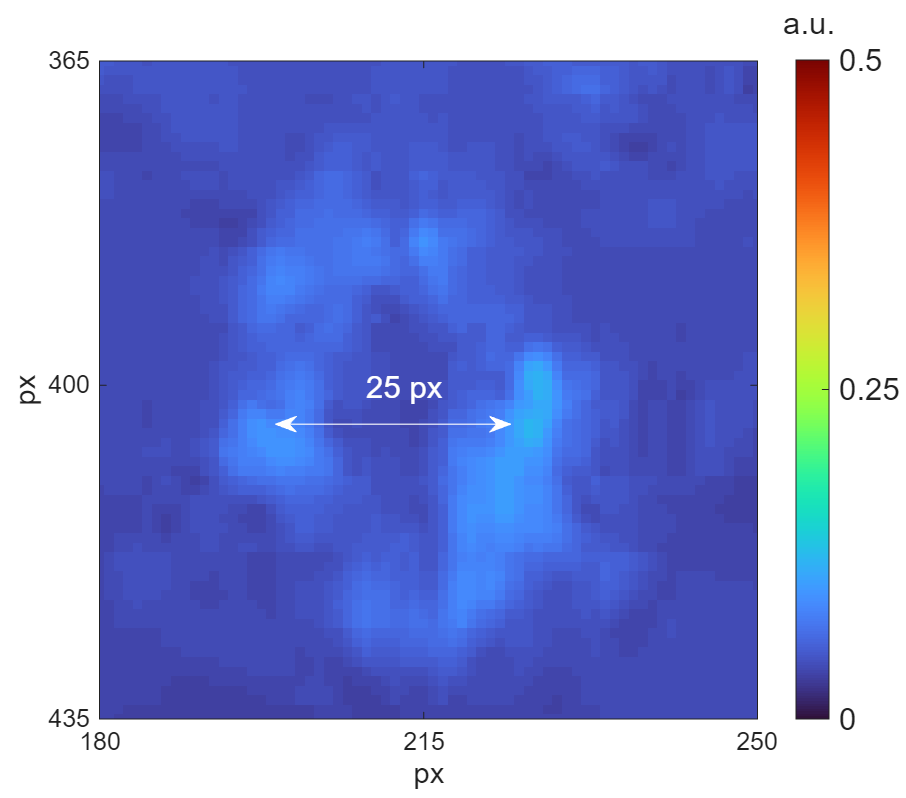} 
    \caption{} 
    \label{fig:lm2} 
  \end{subfigure} 

    \vspace{1ex}
   \begin{subfigure}[b]{0.48\linewidth}
    \centering
    \includegraphics[height=0.2\textheight]{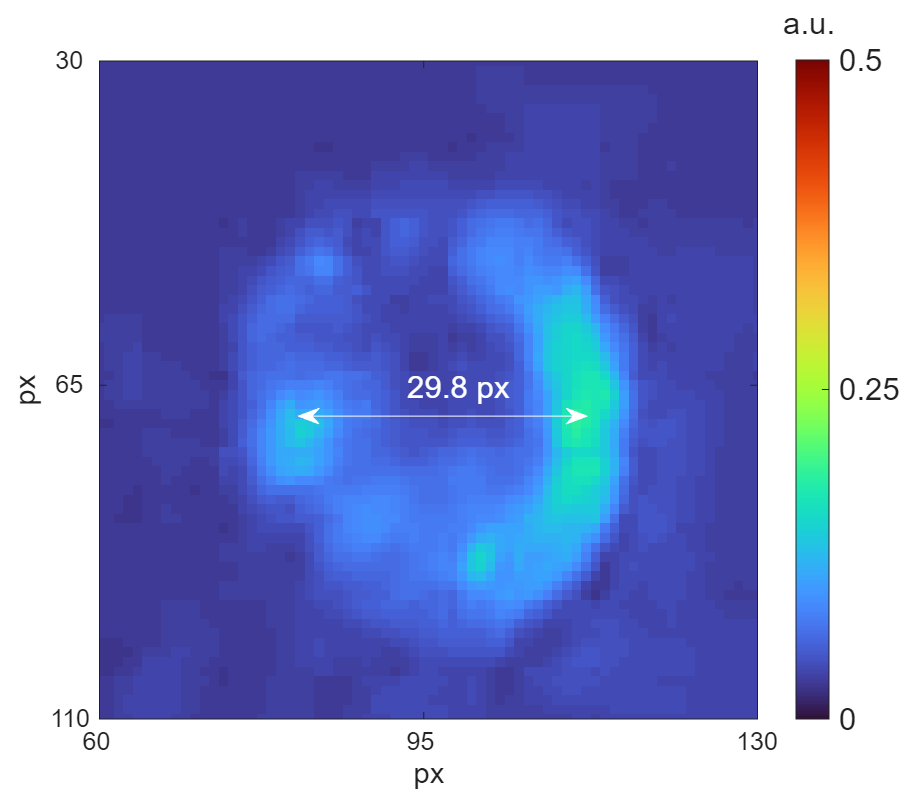} 
    \caption{} 
    \label{fig:l3} 
  \end{subfigure}
  \hspace{\fill}
  \begin{subfigure}[b]{0.48\linewidth}
    \centering
    \includegraphics[height=0.2\textheight]{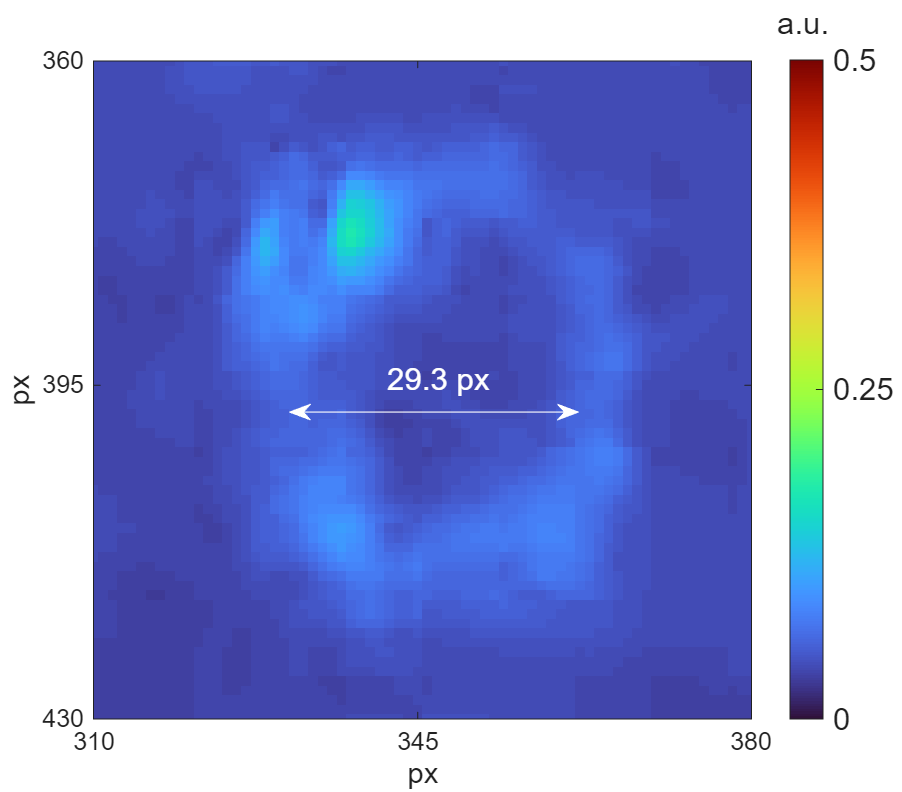} 
    \caption{} 
    \label{fig:lm3} 
  \end{subfigure} 

\caption{Analysis of the vortex beam diameters. In reference to Figure \ref{fig:results_oamarray}, (a) $\ell=1^a$, (b) $\ell=-1^a$, (c) $\ell=1^b$, (d) $\ell=-1^b$, (e) $\ell=2$, (f) $\ell=-2$, (g) $\ell=3$, and (h) $\ell=-3$. The arrow is a guide showing the width of the vortex, and labelled with the width measured in units of pixels. }
\label{fig:analyses_combined} 
\end{figure}

\begin{figure}
    \centering
    \includegraphics[width=0.68\linewidth]{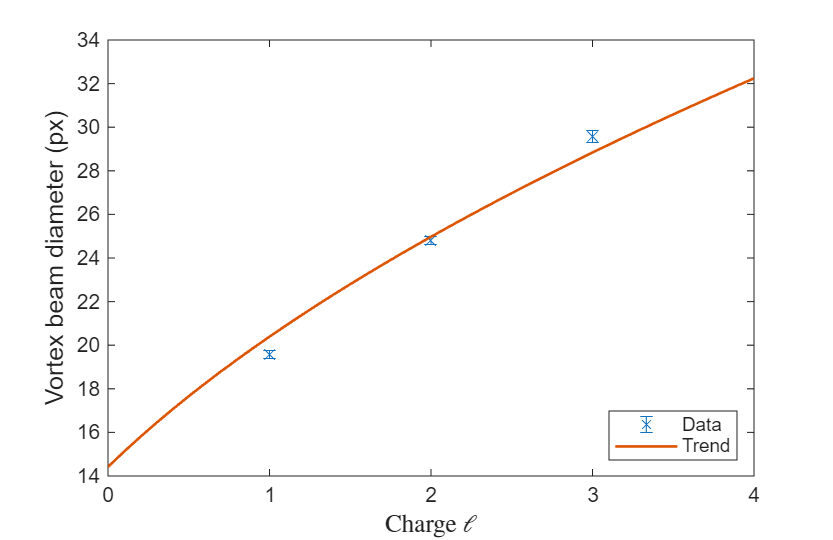} 
    \caption{Variation of the diameter of the vortex beam with topological charge $\ell$ using data from Figure \ref{fig:analyses_combined}. The expression for the trend line fitted to the data is $D \sqrt{\ell + 1}$, where $D$ is the diameter of the Gaussian beam $\ell=0$, and the data shows good agreement with theory. }
    \label{fig:diameterplot}
\end{figure}
% figure proving the charges in the image
%done intensity images, need to arrange and show the relative diameters in a table or in the captions/titles
\newpage

\end{document}